# Growth and Morphology of InN Nanowires on Si⟨111⟩ and Si⟨100⟩ at Back-End-Of-Line Compatible Temperatures

Andrea ORLANDO-CUNNAC [1, 2], Arthur ARNAUD [1], Martien DEN HERTOG [3], Ettore COCCATO [2], Vincent CALVO [2], Jonathan STECKEL [1], Eva MONROY [2]

[1] STMicroelectronics, 850 Rue Jean Monnet, 38920 Crolles, France

[2] Univ. Grenoble-Alpes, CEA, Grenoble INP, IRIG, PHELIQS, 17 av. des Martyrs, 38000 Grenoble, France

[3] Univ. Grenoble-Alpes, CNRS, Grenoble INP, Institut Néel, 43 av. des Martyrs, 38000 Grenoble, France

**Abstract**

InN nanowires were grown on Si⟨111⟩ and Si⟨100⟩ substrates by plasma-assisted molecular beam epitaxy using a thin AlN buffer layer at temperatures compatible with the thermal budget limitation imposed by Back-End-Of-Line processing. Reflection high-energy electron diffraction reveals different nucleation behaviors on the two substrate orientations, with higher structural disorder in the case of Si⟨100⟩. However, vertically aligned nanowires with hexagonal cross section and N polarity are obtained on both substrates. A statistical analysis of nanowire morphology as a function of growth temperature indicates similar trends in diameter, density, and length on Si⟨111⟩ and Si⟨100⟩, which are explained by adatom kinetics during growth. Nanowires on Si⟨100⟩ exhibit improved uniformity and reduced tapering, attributed to the different nanowire nucleation due to microstructural properties of the AlN buffer layer. The results demonstrate the feasibility of growing high-quality InN nanowires on Si⟨100⟩, supporting their potential for monolithic integration of nanowire-based photodetectors on silicon.

## 1. Introduction

III-V semiconductors exhibit interesting properties for optoelectronics, with materials that can cover a wide spectral range, from the ultraviolet to the infrared through varying alloy composition.[1] As a result, III–V materials enable highly efficient emitters and detectors that are difficult to achieve with silicon-based technologies alone. However, the widespread deployment of III-V technologies remains limited by their high production cost and by the difficulty of growing large-diameter III–V wafers comparable to those used in silicon microelectronics. Therefore, the heteroepitaxial integration of III–V materials on silicon have attracted considerable attention as a promising route toward cost-effective and scalable optoelectronic devices.[1]

A major challenge in III-V heteroepitaxy on silicon arises from the significant lattice mismatch.[2] Several approaches have been developed to mitigate this constraint, including pseudomorphic quantum wells,[2] strain-driven Stranski–Krastanov quantum dots,[3] and, more recently, nanowire-based heteroepitaxy. In particular, nanowire growth offers an efficient pathway for elastic strain relaxation through free surfaces, enabled by the high surface-to-volume ratio of the nanowire geometry.

Among III–V materials, InN is of particular interest due to its narrow direct band gap of approximately 0.7 eV and its strong optical absorption, making it a promising candidate for short-wave infrared (SWIR) photodetection.[4,5] In addition, InN can be grown at relatively low temperatures using plasma-assisted molecular beam epitaxy (PAMBE), which is advantageous for integration with silicon-based technologies.[6,7] Previous studies have extensively investigated the growth and morphology of InN nanowires on Si⟨111⟩ substrates,[8–14] demonstrating self-assembled vertically aligned nanowires with well-defined hexagonal cross sections and growth along the c-axis.

Despite these advances, the growth of InN nanowires on Si⟨100⟩ substrates remains unexplored, even though this substrate orientation is predominant in industrial microelectronics. From a technological standpoint, extending InN nanowire growth to Si⟨100⟩ at temperatures compatible with Back End-Of-Line (BEOL) processing—typically below 450 °C—is a critical step toward monolithic integration of InN nanowire-based optoelectronic devices on silicon.

In this work, we present a systematic study of InN nanowire growth on Si⟨111⟩ and Si⟨100⟩ substrates by PAMBE using an AlN buffer layer. The targets are threefold: (i) to assess the reproducibility and temperature-dependent evolution of InN nanowire morphology on Si⟨111⟩, (ii) to demonstrate the feasibility of growing InN nanowires of comparable quality on Si⟨100⟩ substrates, and (iii) to establish growth conditions compatible with BEOL thermal budgets. By combining *in situ* and *ex situ* characterization, we provide insight into the role of substrate orientation and buffer layer nucleation on the nanowire morphology, and identify Si⟨100⟩ as a promising platform for integrated InN nanowire photodetectors.

## 2. Experimental Methods

InN nanowires were grown using PAMBE. A first series of samples were synthesized on ⟨111⟩-oriented silicon wafers, and a second series was grown on ⟨100⟩-oriented silicon, to test the compatibility of the process with the substrates used in imaging technologies. In both cases, the substrates were p-type doped with a resistivity in the range of 0.001-0.005

Ω.cm⁻¹. For both sets of samples, the substrates were precleaned with organic solvents before being introduced in the growth chamber. Then, they were annealed at a temperature of 900° in order to thermally deoxidize the silicon surface. Deoxidation is confirmed by the observation of the 7×7 reconstruction of the silicon surface using reflection high-energy electron diffraction (RHEED).

Next step involved the deposition of a two-step AlN buffer layer, which is known to promote vertical growth in the case of GaN nanowires.[15] The active N flux was fixed to obtain a growth rate of 0.5 monolayers per second (ML/s) under N-limited conditions, and the Al cell temperature corresponds to the stoichiometric flux (Al/N ratio = 1.00±0.05). In a first stage, the substrate was cooled down to 350°C and 1.2 nm of AlN were deposited by alternation of Al and N (two 5 s/5 s cycles). The substrate was then warmed up to the nanowire growth temperature (between 363°C and 393°C, depending on the sample). The Al and N cells were then opened simultaneously to complete an AlN buffer layer with a total thickness of 5 nm. The InN growth was performed with an In/N ratio = 0.15±0.01), with both In and N cells were opened simultaneously for 3 hours. Growth was monitored continuously using RHEED. Note that the substrate temperature in the growth chamber was calibrated based on the eutectic point of AuGe at 361°C.[16]

The evolution of nanowire morphology was analyzed in a ZEISS_ULTRA55A field-emission scanning electron microscope (SEM). Nanowire diameters were automatically extracted from the SEM images using the ImageJ algorithm, with an average calculated from approximately 1,000 nanowires. The software was also used to determine the average space occupied by the nanowires in each top-view image. The nanowire density was derived from these two parameters. The nanowire length and taper ratio were extracted manually from cross-section SEM images, by calculating the average of around 30 wires. The taper ratio is defined as the ratio between the nanowire diameter at the top and at the bottom of the nanowire, providing a quantitative measurement of the nanowire tapering. Higher taper ratio values indicate a more pronounced tapering effect.

For scanning transmission electron microscopy (STEM) characterization nanowires were dry dispersed on holey carbon grids. High angle annular dark field (HAADF) STEM studies were performed using a JEOL Neo Arm working at 200 kV at about 27.5 mrad convergence angle equipped with a probe aberration corrector.

### 3. Results

Figure 1 shows RHEED patterns recorded at different stages of growth for the two substrate orientations under study. Figures 1a and b show the diffraction patterns after the deposition of 5 nm of AlN. In the case of Si⟨111⟩ (Figure 1(a)), we observe a spotty diffraction pattern with additional reflections, which is an indication of the formation of crystalline islands. A ring-like pattern around the main diffraction spots indicates a certain degree of misorientation between the islands. In the case of growth on Si⟨100⟩, the diffraction spots disappear and the pattern consists of clearly resolved rings, which points to randomly oriented nucleation.

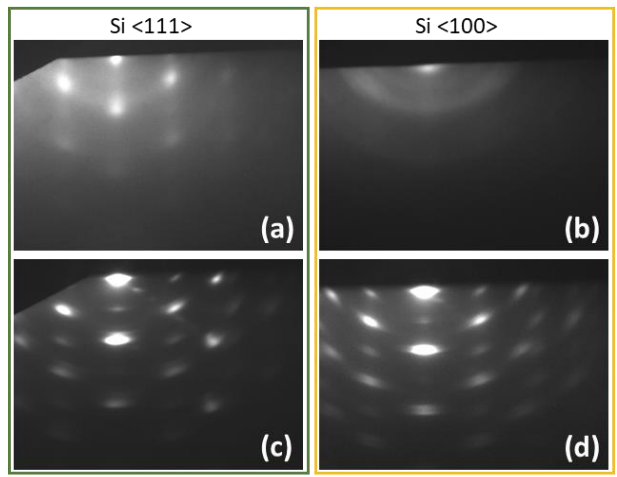

*Figure 1 : RHEED patterns after completing the 5-nm-thick AlN buffer layer (a) on Si <111> and (b) on Si <100>, and after 3 hours of InN growth (c) on Si <111> and (d) on Si <100>.*

Figures 1(c) and (d) show the RHEED pattern after 3 hours of growth of InN on Si⟨111⟩ and ⟨100⟩, respectively. The images are very similar, with clearly resolved diffraction spots, as expected in the case of nanowire growth. The elongation of the spots following a ring pattern indicates a certain degree of tilt of the wires in both cases. Based on these observations, we can speculate that the nanowire morphology might be similar for Si⟨111⟩ and ⟨100⟩, in spite of the different quality of the nucleation layer.[17,18]

The morphology of the resulting InN nanowires is illustrated by SEM images in Figure 2, where Figs. 2(a-c) correspond to nanowires grown on Si⟨111⟩ and Figs. 2(d-e) describe nanowires on Si⟨100⟩. Regardless of the substrate orientation, the nanowires are well-defined, exhibit hexagonal cross section, and have similar overall dimensions.

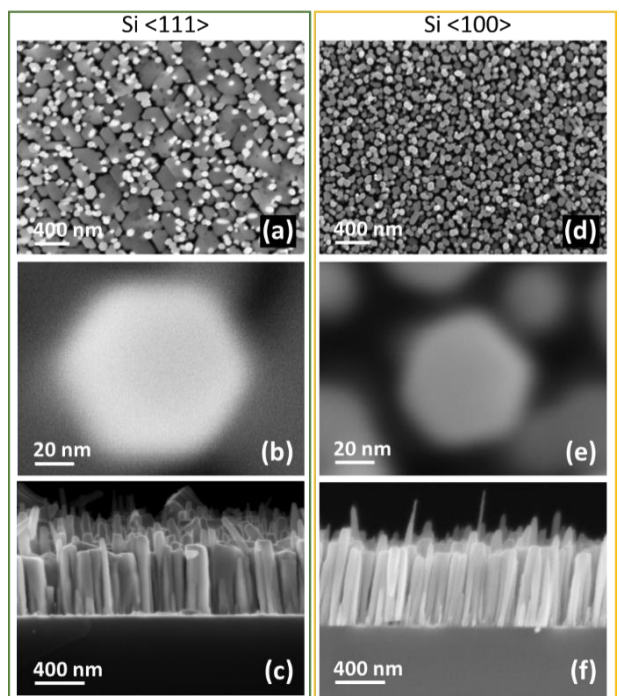

*Figure 2: (a-c) SEM images of InN nanowires grown on Si<111>: (a) top view, (b) zoomed view of one nanowire showing the hexagonal cross section, and (c) cross-section view. (a-c) SEM images of InN nanowires grown on Si<100>: Equivalent images of InN nanowires grown on Si <100>.*

Figure 3 presents a comparative statistical study of nanowires gown on Si⟨111⟩ and ⟨100⟩ substrates, based on various morphological parameters of the nanowires extracted from SEM images as explained in the description of experimental methods. Error bars represent the standard deviation of the measurements.

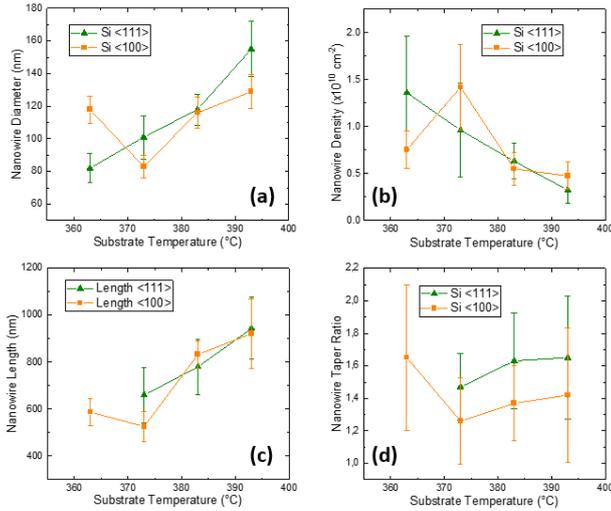

*Figure 3: Evolution of the (a) average nanowire diameter, (b) nanowire density, (c) average nanowire length, and (d) taper ratio as a function of the growth temperature. Each panel compares the results obtained on Si <111> (red) and <100> (orange) substrates.*

Figure 3(a) shows the evolution of the average nanowire diameter as a function of growth temperature. For both Si⟨111⟩ and ⟨100⟩ substrates, the diameter follows the same general trend, increasing with increasing temperature. For growth temperatures above 380°C, nanowires grown on Si⟨100⟩ substrates exhibit slightly smaller diameters than those on ⟨111⟩ substrates.

Figure 3(b) presents the evolution of nanowire density as a function of substrate temperature. No significant difference in density is observed between the two substrates. In both cases, the density decreases as the growth temperature increases. Figure 3c describes the evolution of the average nanowire length as a function of growth temperature. A clear, similar trend is observed for both substrates, with the nanowire length increasing as the temperature rises. Finally, Figure 3d illustrates the evolution of the nanowire taper ratio as a function of growth temperature. In view of the next fabrication steps towards InN nanowire-based photodetectors, which involve passivating the nanowire interfaces, it is desirable that the nanowires exhibit a taper ratio as close to 1 as possible. For both substrate orientations, the error bars are too large to extract a trend as a function of the growth temperature. However, it is clearly observed that nanowires grown on Si⟨100⟩ exhibit less tapering than those on Si⟨111⟩.

Note that error bars are larger for parameters extracted from cross-sectional images, namely the taper ratio and length (Figures 3c and 3d). This is due to the fact that the analysis considered fewer nanowires compared to top-view images.

The polarity of the nanowires has been investigated using STEM images, as illustrated in Fig. 4 for specimens grown on (a-d) Si⟨111⟩ and (e-h) Si⟨100⟩. Intensity profiles extracted from high-resolution HAADF STEM images allow us to determine the relative position of In and N atomic columns, confirming their presence in both cases at least part of the nanowires exhibit nitrogen polarity, i.e., they grow along the [000-1] crystallographic axis.

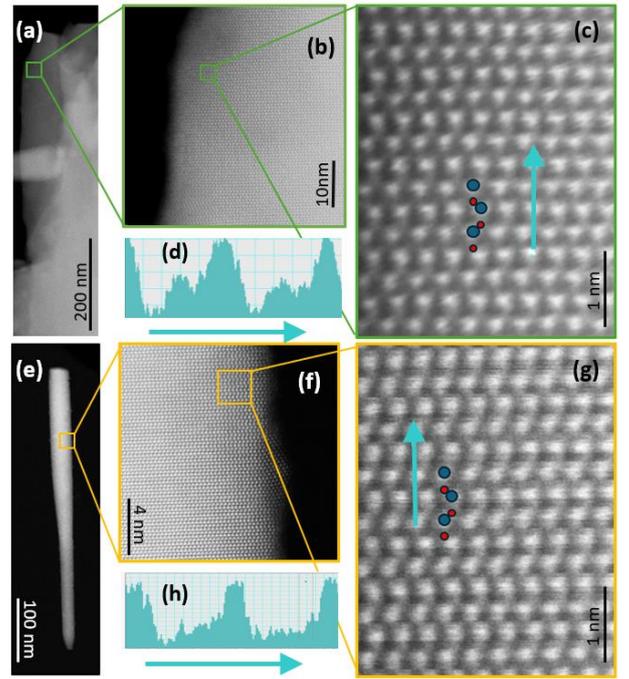

*Figure 4: Structural and polarity analysis of InN nanowires by STEM: (a,e) Low-magnification STEM images of representative InN nanowires, with arrows indicating the growth direction. (b,f) Higher-magnification views of the nanowire sidewalls; the boxed areas indicate the regions used for atomic-resolution imaging. (c,g) Atomic-resolution STEM images revealing the wurtzite crystal structure, with the projected positions of In (blue) and N (red) atomic columns overlaid, allowing determination of the nanowire polarity. The arrows indicate the growth direction, confirming nitrogen polarity. (d,h) Intensity profiles extracted along the arrow in in (c) and (g), respectively, showing periodic contrast consistent with the atomic column arrangement.*

**4. Discussion**

With a view towards the fabrication of integrated InN nanowire photodetectors with CMOS, the targets of this work was to conduct an in-depth study of the morphological parameters of InN nanowires as a function of growth temperature on Si⟨111⟩, to assess the feasibility of high-quality InN nanowire growth on Si⟨100⟩, to align with existing silicon imaging technologies on the market, and to evaluate if such nanowires can be grown at temperatures compatible with the thermal budget limitations imposed by Back End Of Line (BEOL) processes.[19]

It was important to determine the substrate temperature during growth precisely. Accurate temperature measurement in MBE is challenging because the vacuum environment in the growth chamber limits thermal conduction, and temperatures around 400°C are difficult to measure reliably using a pyrometer due to background radiation. In this study, the temperature was monitored using a thermocouple that was not in direct contact with the sample, resulting in a significant offset between the real and measured temperature.

To obtain a more accurate estimate of the actual growth temperature, we followed a calibration procedure based on the use of a material that exhibits a well-defined phase transition observable by RHEED at a temperature close to the growth conditions. For this purpose, we chose AuGe because it has a eutectic temperature of 361°C.[16] All the growth temperatures reported in this manuscript were corrected taking this eutectic measurement into account.

If we focus on the growth of nanowires on Si⟨111⟩, the observation of a high density of self-assembled vertically aligned nanowires with a hexagonal cross section is consistent with previous reports.[14] The trends observed in

Figure 3 are in good agreement with those reported by A.T.M. Golam Sarwar et al.[8], namely an increase in the nanowire diameter, taper ratio, and length, accompanied by a decrease in nanowire density as the growth temperature increases. These four trends are interrelated and can be understood within the framework of existing models describing the behavior of adatoms during nanowire growth in GaN nanowires.[20,21] In particular, these models account for the diffusion of group-III metal adatoms along the nanowire sidewalls, where they react with group-V (N) adatoms, contribution to both axial and radial growth. As the growth temperature increases, the diffusion length of the adatoms also increases. As a direct consequence, adatoms are more likely to aggregate and form larger nucleation sites. Since the total amount of material available is the same regardless of the substrate temperature neglecting re-evaporation, these larger nucleation sites necessarily result in a reduced number of nanowires. This explains the opposite trends observed for nanowire diameter, in Fig. 3(a), and density, in Fig. 3(b), when varying the growth temperature. The same mechanism also accounts for the increase in nanowire length with increasing growth temperature: with fewer wires competing for the same amount of material, a larger fraction of the atom flux contributes to axial growth, resulting in longer nanowires, as observed in Fig. 3(c).

Turning now to growth on Si⟨100⟩ substrates, Figs. 1(d) and 2(d,e,f) clearly demonstrate the successful formation of InN nanowires in terms of density, shape, and vertical alignment. This enabled us to conduct a comparative statistical analysis of the morphological parameters obtained on Si⟨111⟩ and ⟨100⟩ substrates (Fig. 3). Overall, the nanowire morphology follows the same trends as a function of the growth temperature for both substrate orientations. However, systematic differences are observed when comparing the two cases. First, nanowires on Si⟨100⟩ exhibit a smaller average diameter and much better uniformity in diameter distribution (lower standard deviation), as shown in Fig. 3(a). Figure 3(c) also points to a more uniform distribution of nanowire lengths on Si⟨100⟩, which adds up to the reduced tapering observed in Fig. 3(d).

Comparing the evolution of the morphological parameters of InN nanowires grown on Si⟨111⟩ and ⟨100⟩ substrates reveal systematic differences that cannot be attributed solely to intrinsic adatom diffusion on the nanowire sidewalls, since in both cases the nanowires grow along the (000-1) direction and expose identical m-plane {1-100} sidewalls. Instead, the observed morphology can be more consistently explained by differences in nucleation conditions imposed by the AlN buffer layer. On Si⟨111⟩, the AlN buffer exhibits a higher degree of epitaxy, leading to a smaller number of preferential nucleation sites. This results in stronger wire–wire competition during growth, which promotes broader distributions in nanowire diameter and length.

In contrast, on Si⟨100⟩ the AlN buffer exhibits significant in-plane twist, indicative of a higher density of grain boundaries and structural disorder. This microstructural complexity favors a more spatially homogeneous and denser nucleation of nanowires, leading to narrower distributions in diameter and length. In a denser nanowire ensemble, geometric effects such as mutual shadowing and reduced flux capture by individual nanowire tips can limit preferential radial growth near the top of the nanowires, thereby reducing tapering.

In view of the results presented in this work, InN nanowires grown on Si ⟨100⟩ emerge as a more suitable platform for integrated photodetectors than those grown on Si ⟨111⟩. Their improved morphological homogeneity and reduced tapering significantly simplify subsequent processing steps, including surface passivation, planarization, and electrical contacting. In addition, we demonstrate that high-quality InN nanowires can be grown at temperatures compatible with BEOL thermal budgets, further supporting their potential for monolithic integration of InN nanowire photodetectors on silicon CMOS.

**5.Conclusion**

In conclusion, we have conducted a systematic comparative study of InN nanowire growth on Si⟨111⟩ and Si⟨100⟩ substrates as a function of growth temperature using plasma-assisted molecular beam epitaxy. Despite markedly different nucleation behaviors of the AlN buffer layer on the two substrate orientations, high-quality, vertically aligned InN nanowires displaying nitrogen polarity were successfully grown in both cases. Statistical analysis reveals that, while the overall temperature-dependent trends in nanowire diameter, density, and length are similar, nanowires grown on Si⟨100⟩ offer clear advantages for integration due to their improved morphological homogeneity and reduced tapering compared to those on Si⟨111⟩. These differences are attributed to the higher density and spatial uniformity of nucleation sites associated with the structurally disordered AlN buffer on Si⟨100⟩. Importantly, the nanowires can be grown at temperatures compatible with BEOL thermal budgets. These results demonstrate that Si⟨100⟩ substrates constitute a particularly promising platform for the monolithic integration of InN nanowire-based photodetectors with silicon CMOS technologies.

The TEM facility JEOL NEOARM was co-financed by the European Union under the European Regional Development Fund (ERDF) (Contract No. RA0023813). A CC-BY public copyright license has been applied by the authors to the present document and will be applied to all subsequent versions up to the Author Accepted Manuscript arising from this submission, in accordance with the grant's open access conditions.

The data that support the findings of this study are available on request from the corresponding author. The data are not publicly available due to privacy restrictions.